\documentstyle[prl,twocolumn,floats,epsf,aps]{revtex}
\def\ivpb{\ignorespaces{\parfillskip=0pt\par\eject}
          \noindent\ignorespaces}
\def\hb{\hfill\break}
\belowdisplayskip \abovedisplayskip
\abovedisplayshortskip  \abovedisplayskip
\belowdisplayshortskip \abovedisplayskip

\begin{document}
\draft
\title{Contrasts between coarsening and relaxational dynamics of surfaces}
\author{Martin Siegert, Michael Plischke}
\address{Physics Department, Simon Fraser University,
Burnaby, British Columbia, Canada V5A 1S6}
\author{R.K.P. Zia}
\address{Center for Stochastic Processes in Science and Engineering,\\
Physics Department, Virginia Polytechnic Institute and State University,
Blacksburg, VA 24061-0435}
\author{\vskip -2\baselineskip\small(February 10, 1997)\break}
\author{\vbox{\hsize 14cm\small%
\hbox to\hsize{\quad We discuss static and dynamic fluctuations
of domain walls separating areas of constant but}
different slopes in steady-state configurations of
crystalline surfaces both by an analytic treatment of the appropriate
Langevin equation and by numerical simulations. In contrast to other
situations that describe the dynamics in Ising-like systems such as models A
and B, we find that the dynamic exponent $z=2$ that governs the domain wall
relaxation function is {\em not} equal to the inverse of the exponent
$n\approx 1/4$ that describes the coarsening process that leads to the steady
state.
\hb\hb
\leftline{PACS numbers: 64.60.Cn, 05.70.Ln, 64.60.My, 68.35.Ja}}}
\address{\vskip -0.5\baselineskip}
\maketitle
\title{Contrasts between coarsening and relaxational dynamics of surfaces}
\author{Martin Siegert, Michael Plischke}
\address{Physics Department, Simon Fraser University,
Burnaby, British Columbia, Canada V5A 1S6}
\author{R.K.P. Zia}
\address{Center for Stochastic Processes in Science and Engineering,\\
Physics Department, Virginia Polytechnic Institute and State University,
Blacksburg, VA 24061-0435}
\date{\today}
\maketitle
The description of a number of surface relaxation processes in which the
primary relaxation mechanism is diffusion of particles along the interface
begins, in a continuum picture, with a Langevin equation of the form 
\begin{equation}
\partial_th({\bf r},t)+\nabla {\bf \cdot j}[h({\bf r},t)]=\eta({\bf r},t)
\label{conserv}
\end{equation}
where ${\bf r}=(x_1,x_2)$ denotes the co-ordinates in a reference plane, $h$
is the height of the surface above it, ${\bf j}$ represents a surface
diffusion current, and $\eta$ accounts for thermal noise. The conservation
law in (\ref{conserv}) reflects the assumed absence of voids, overhangs and
evaporation of particles. For example, this equation was applied to the case
of spinodal decomposition of an unstable crystal facet\cite{nigel,liu}. In
that case, $\nabla\cdot{\bf j}$ is simply the functional derivative of the
surface free energy with respect to the function $h$.

Our interest here is another case: the growth of films by molecular beam
epitaxy (MBE). It was realized some time ago\cite{villain,KPS} that in MBE
the breaking of detailed balance by the deposition process and the presence
of step-edge barriers could lead to a nonequilibrium diffusion current that
destabilizes a singular surface and leads to the formation of mounds or
pyramids\cite{orme,SP1}. This behavior is observed in simulations of
discrete microscopic models as well as numerical integration of the
following version of Eq.~(\ref{conserv}): 
\begin{eqnarray}
\partial_th({\bf r},t)&=&-D\Delta\Delta h({\bf r},t)-v\Delta h({\bf r},t) 
\nonumber \\
&&+u\sum_\nu\partial_\nu\left[\partial_\nu h({\bf r},t)\right]^3
+\eta({\bf r},t)\ ,  \label{current}
\end{eqnarray}
where $\partial_\nu\equiv\partial/\partial x_\nu $, $\nu =1,2$, and 
$\Delta\equiv\partial_1^2+\partial_2^2$. The first term on the right arises
from the equilibrium diffusion current\cite{mullins}, the next two represent
the nonequilibrium diffusion current which with $v,\,u>0$ destabilize the
singular surface ($\partial_\nu h=0$) and lead to a steady state composed of
facets of slope $\partial_\nu h=\pm\sqrt{v/u}$. In particular, during the
late stages, both methods reveal that these three-dimensional features
coarsen as function of time with the characteristic length scale $R$
evolving as $t^n$ with $0.16\lesssim n\lesssim0.25$. At present, this
property is not 
\ivpb
well understood.

In this letter, we focus on another aspect of the dynamics of this model,
namely, relaxation into a steady state consisting of two domains. We also
investigate an intimately related subject: the fluctuations of the interface
separating these domains. Here, a domain refers to an area of constant slope
mentioned above. For the analytic part, we will rely on Eq.~(\ref{current}),
setting, for simplicity, $u=v=2D=1$. To compare with these results we will
present data from both simulations and numerical integration of 
(\ref{current}). Our main conclusion is that $z$, the exponent associated with
relaxational dynamics, is 2. Specifically, we find that the relaxation time
of a domain wall fluctuation of wavevector $k$ scales as $\tau_k\sim k^{-2}$.
Thus, we have a system which violates $z=1/n$, a relation considered to be
generic, coming from the dynamic scaling assumption with only a single time
scale. Two well-studied systems obeying this relation are models A and 
B\cite{HH}. In the former, the Allen-Cahn equation\cite{AC} used to describe
coarsening dynamics is even identical to the equation which describes the
slowest relaxational mode\cite{BDJZ}, giving $z=1/n=2$. Similarly, the
dynamics of model B is known to display a coarsening exponent 
$n=\frac13$\cite{rogers} and a relaxation exponent $z=3$\cite{bausch}.
In contrast to this simple situation, the behavior of our system is
considerably richer and more intriguing. Not only do we find $z=2\neq 1/n$,
we also observe the steady-state structure factor to scale as
$S(k)\sim k^{-2}$. These results indicate that the coarsening process is
controlled by defects that are not present in the stationary state.

It is tempting to compare our system with model B, both having similar
non-linearities in the fields and conservation laws. However, there are
serious differences, which are most apparent if we rewrite (\ref{current})
in terms of a two-component ``order parameter'', $m_\nu =\partial _\nu h$: 
\begin{eqnarray}
\partial _tm_\nu ({\bf r},t) 
&=&-\hbox{$1\over2$}\Delta\Delta m_\nu({\bf r},t)-\Delta m_\nu({\bf r},t)
\nonumber \\
&&+\sum_\mu\partial_\nu \partial_\mu\left[m_\mu^3({\bf r},t)\right]
+\partial_\nu\eta({\bf r},t)\ ,  \label{langm}
\end{eqnarray}
with the constraint $\partial_\mu m_\nu =\partial_\nu m_\mu $. By
contrast, for a model B with two decoupled order parameters and a simple
free energy density 
${\cal F}=\sum_\nu\frac14(\nabla m_\nu )^2-\frac12m_\nu^2+\frac14m_\nu^4$,
the third term would read $\Delta m_\nu^3$ and no constraint would be imposed
on $m_\mu$.

We begin by discussing a stationary state with two domain walls. Such a
state can be imposed on a system with substrate dimensions $L_x$, $L_y$ by
screw boundary conditions in $x$: $h(x+L_x,y)=h(x,y)+L_x$ and periodic ones
in $y$: $h(x,y+L_y)=h(x,y)$. Thus the average slope in $x$ can be unity,
which is one of the preferred ground states. In the $y$-direction, there is
a single ridge and a single valley and, except for two regions of
characteristic size $\xi\approx\sqrt{2D/v}=1$, $\partial_yh(x,y)=\pm 1$.
We denote the steady-state height function by $h_c(x,y)=x+\Phi_c(y)$ and
its derivative with respect to $y$ by 
$\varphi_c(y)\equiv\partial_yh_c(x,y)=\Phi_c'(y)$. Though $\Phi_c$
can be found exactly, its details will turn out to be unimportant. To provide
the reader some familiar ground, we quote an example in the limit 
$\xi /L_y\to 0$: $\varphi_c(y)\to\tanh ((y-l)/\xi )-\tanh ((y-3l)/\xi )-1$,
with $l=L_y/4$, where the valley and ridge are located at $y=l$ and $y=3l$,
respectively.

Assuming now that $h({\bf r},t)=h_c({\bf r})+\chi({\bf r},t)$ and
linearizing (\ref{current}), we find that $\chi$ satisfies the Langevin
equation 
\begin{equation}
\partial_t\chi({\bf r},t)=-{\cal L}\chi({\bf r},t)+\eta({\bf r},t)
\label{linchi}
\end{equation}
with the Hermitian operator 
${\cal L}=\frac12\Delta\Delta
-\partial_x[V(\partial_xh_c)\partial_x]-\partial_y[V(\partial_yh_c)\partial_y]$
and $V(\zeta )=3\zeta^2-1$. The fluctuations of the surface can thus be
expanded 
\begin{equation}
\chi ({\bf r},t)=\sum_\ell A_\ell(t)v_\ell({\bf r})\ ,  \label{chi_expand}
\end{equation}
in terms of the eigenfunctions $v_\ell$ which solve the eigenvalue problem 
\begin{equation}
{\cal L}v_\ell({\bf r})=\lambda_\ell v_\ell({\bf r})
=\lambda_n(q)\psi_n(q,y)e^{-iqx}\ .  \label{eigenv}
\end{equation}
Note that, for the last equality, we have exploited the translational
invariance in $x$.

There are several branches of the eigenvalue spectrum. Two of these, the
height fluctuation modes and the `Goldstone' branch, terminate at $q=0$ with
eigenvalue zero. The eigenfunctions of these zero-frequency modes are simply 
$\psi_H(0,y)=1$ and $\psi_G(0,y)=\varphi_c(y)$. The result $\lambda_H(q=0)=0$
follows from translational invariance in $h$, while $\lambda_G(q=0)=0$ is due
to Goldstone's theorem. The higher modes of the H-branch are simply obtained: 
$\psi_H(q,y)=\exp (iqx)$ with $\lambda_H(q)=2q^2+q^4/2$. The nontrivial
G-modes, on the other hand cannot be determined exactly. Focusing on small $q$,
we write ${\cal L}={\cal L}_0+{\cal V}(q)$ with 
${\cal V}(q)=q^2(2-\partial_y^2+q^2)$ as a perturbation. Perturbation theory
or the variational method 
\begin{equation}
\lambda \le \int dy\,\varphi {\cal L}\varphi \biggl/\int dy\,\varphi^2
\label{lambda_var}
\end{equation}
with a trial function $\varphi =\varphi_c$ both yield $\lambda_G(q)
=2q^2+{\cal O}(q^2/L_y,q^4)$.
\eject
Finally, a third set of low-lying modes can be identified --- the breathing
modes in which the ridge and valley move in opposite directions. An
appropriate choice of the variational eigenfunction in (\ref{lambda_var})
for the breathing mode is $\psi_B(0,y)\simeq \tanh(y-l)+\tanh(y-3l)-y/l+2$
where the last two terms guarantee that the integral of this function is
zero and that periodic boundary conditions are satisfied. A simple
calculation using $\psi_B'(y)\simeq |\psi_G'(y)|-1/l$ then
produces the bound $\lambda_B(q=0)\leq 96/L_y^2$. It is also easy to see
that higher modes have the property 
$\lambda_B(q)=c/L_y^2+2q^2+{\cal O}(q^2/L_y,q^4)$ with some numerical
constant $c\le 96$. Thus, the breathing mode has a gap $\sim 1/L_y^2$.
However, for quadratic systems ($L_x=L_y$) this gap is of the same order
as $q^2$ and therefore $\lambda_B$ is of the same order as $\lambda_G$ for
nonzero $q$. Thus, all eigenvalues that we have found so far scale in lowest
order as $\lambda_n(q)\sim q^2$. For an exactly solvable version of this
model\cite{ZSP} it is found that this statement is indeed correct for
{\it all\/} eigenvalues. Perhaps $\lambda\sim q^2$ should not be a surprise,
considering the presence of at least two gradients on the RHS
of (\ref{current}) or (\ref{langm}). This is also the case, for a
{\em homogeneous} state, in model B. The contrast with model B becomes
dramatic only when we consider {\em inhomogeneous} states, which admit
Goldstone modes. In model B, the fact that ${\cal L}$ is not
Hermitian leads to $\lambda\sim q^3$\cite{bausch}
while here $\lambda\sim q^2$.

We now can calculate two-point correlation functions such as
$S({\bf k,}\tau)\equiv \int d^2r'd^2r\,e^{i{\bf k\cdot r}}
\left\langle\chi({\bf r}',t)\chi({\bf r}'+{\bf r},t+\tau )\right\rangle$.
Expanding the noise $\eta $ in terms of the eigenfunctions, we write 
$\eta({\bf r},t)=L_x^{-1/2}\sum_{n,q}B_n(q,t)\psi_n(y)e^{-iqx}$. Using
Eqs.~(\ref{linchi}), (\ref{chi_expand}) and
$\langle\eta({\bf r},t)\eta({\bf r}',t)\rangle
=2\epsilon\delta({\bf r}-{\bf r}')\delta(t-t')$, we find 
\begin{equation}
S({\bf k},\tau )=\sum_n{\frac{\epsilon}{\lambda_n(q)}}\widehat{\psi}_n(p) 
\widehat{\psi }_n(-p)e^{-\lambda_n(q)\tau }\ ,  \label{Skt}
\end{equation}
where ${\bf k}\equiv(q,p)$ and $\widehat{\psi}_n(p)$ is the Fourier
transform of the (normalized) $\psi_n(y)$. This is a general result, showing
that the term associated with the lowest eigenvalue will dominate the sum as 
$\tau \rightarrow \infty $. However, in our case, there is no gap, i.e., all 
$\lambda$'s are ${\cal O}(1/L^2)$, so that there is no valid reason to keep
only one or two terms. Because of the dependence on the Fourier transforms
of the eigenfunctions, $S(q,p,\tau)$ will not display simple scaling, apart
from the special case $p=0$. However, since all eigenvalues are proportional
to $k^2$ the time-time correlation function 
$\Phi({\bf k},\tau )=S({\bf k},\tau )/S({\bf k},0)$ depends only on $k^2\tau$
and thus the dynamic exponent that governs the fluctuations in the steady
state is $z=2$.

We are also interested in the interface structure factor $S_I(q,\tau)$ as
it may provide more insight into the coarsening behavior. The position of
the interfaces is determined by the condition $\partial_yh(x,y,t)=0$. This
equation has two solutions $y_1(x,t)$ and $y_2(x,t)$ that correspond to the
positions of the top and the valley of the height profile. The interface
structure factor can be calculated using either of the two functions, 
$S_I(q,t)=\langle \widehat{y}_1(q,t+\tau )\widehat{y}_1(-q,t)\rangle
=\langle\widehat{y}_2(q,t+\tau )\widehat{y}_2(-q,t)\rangle $. In linear
approximation we find that 
$y_1(x,t)=-\partial_y\chi({\bf r},t)/\partial_y^2h_c(y)|_{y=l}$, and therefore 
\begin{eqnarray}
S_I(q,\tau)&=&{1\over[h''_c(l)]^2}
\langle\widehat{\partial_y\chi }|_{y=l}(q,t+\tau )
\widehat{\partial_y\chi|}_{y=l}(-q,t)\rangle   \nonumber \\
&=&{1\over[h''_c(l)]^2}\sum_n[\partial_y\psi_n(l)]^2
{\epsilon\over\lambda_n(q)}e^{-\lambda_n(q)\tau}\ .  \label{Skt_i}
\end{eqnarray}
Because of the symmetries of the steady-state configurations all
eigenfunctions are either symmetric or antisymmetric about $y=0$ and $y=l$.
Of these four classes, only eigenfunctions that have a nonzero derivative at 
$y=l,3l$ contribute to $S_I(q,t)$ (\ref{Skt_i}). The most important modes
with this property are the Goldstone mode $\psi_G(y)$ and the breathing
mode $\psi_B(y)$. Since 
$\partial_y\psi_G|_{y=l},\partial_y\psi_B|_{y=l}\sim L_y^{-1/2}$ we therefore
find 
\begin{equation}
S_I(q,\tau )
={\epsilon\over L_y}\left[{c_G\over\lambda_G(q)}e^{-\lambda_G(q)\tau }
+{c_B\over\lambda_B(q)}e^{-\lambda_B(q)\tau}+\ldots\right]\ , \label{Si_res}
\end{equation}
where $c_G$ and $c_B$ are numeric constants. Hence the interfaces
fluctuations are governed by a dynamic exponent $z=2$ as well. Furthermore,
Eq.~(\ref{Si_res}) shows that the interface structure factor scales as 
$S_I(q)\sim 1/(q^2L_y)$. A major consequence is that the interface width, 
$w(t)=[L_x^{-1}\int dx\,y_1^2(x)]^{1/2}\sim [L_x/L_y]^{1/2}$, does {\em not}
diverge in the thermodynamic limit (provided $L_x\sim L_y$). Hence the
pyramids that form in the coarsening dynamics are separated by {\it flat\/}
interfaces in striking contrast to the interfaces in models A or B, in which
the widths diverge as $\sqrt{L}$.

The coarsening dynamics of Eqs.~(\ref{conserv}) and (\ref{langm}) are still
not well understood. Recently, Rost and Krug\cite{RK} presented a scaling
theory of coarsening for the rotationally invariant current 
${\bf j}_{\rm iso}={\bf m}(m_0^2-m^2)$ that neglects the crystalline symmetries
of the growing film. It is questionable whether such scaling theories are
correct. The growth law for the average domain size depends usually on
{\it two\/} lengthscales: the average domain size $R(t)$ itself and the width
of the domain walls $\xi$. This was made clear by Bray and Rutenberg\cite{BR}.
For the Cahn-Hilliard equation $\partial_tm=-\Delta[\Delta m+m(1-m^2)]$ they
find that the domains grow as $\dot{R}(t)\sim 1/[\xi R^2(t)]$ leading to a
coarsening exponent $n=1/3$. Scaling and mean-field-like theories\cite{MVZ}
for the Cahn-Hilliard equation yield $n=1/4$. This result is obtained, if
the assumption is made that there is only one relevant lengthscale, the
average domain size, so that $\xi$ is replaced by $R(t)$ in the expression
above. Since the dimensionality of the operators in Eq.~(\ref{langm}) is the
same as those of the Cahn-Hilliard equation, scaling theories and
dimensional analysis for equation (\ref{conserv}) with a current 
${\bf j}\ne{\bf m}f(|{\bf m}|)$ are not very convincing. For the case of an
isotropic current as studied in Ref.~\cite{RK} the situation seems to be not
so bad: At least in the analogous case of the Cahn-Hilliard equation only
logarithmic corrections to the growth law are found\cite{BR} if compared
with scaling theories.

Rost and Krug\cite{RK} assume that the average slope approaches the limiting
value $m_0$ in a power law manner, $\langle m\rangle=m_0-ct^{-\alpha'}$, and
they find that the exponent $\alpha' $ is linked to the coarsening exponent
through $\alpha'=2n$. For model B the corresponding relation is different;
the order parameter approaches its limiting value as $m_0-m(t)\sim t^{-1/3}$.
Nevertheless, our calculations do lend some support for the relation 
$\alpha'=2n$ even in the case of an anisotropic surface current. A central
result of our study is that all modes are gapless, $\lambda({\bf k})\sim k^2$.
Consequently, slope fluctuations decay like 
$(\delta m(t))^2\sim\int d^2r[\partial_x\chi({\bf r},t)]^2\sim t^{-1/2}$
and therefore $\alpha'=1/2$. Hence the relation $\alpha'=2n$ yields $n=1/4$
in agreement with numerical solutions\cite{SP1,S_cam} of Eq.~(\ref{current}).
Therefore, the differences between model B and surface dynamics that
became apparent in the calculations presented above, in particular the fact
that $n\ne z^{-1}$, may also be responsible for the difference in the growth
law.

\begin{figure}[b]
\centerline{\epsfxsize=7cm\epsfbox[65 196 526 535]{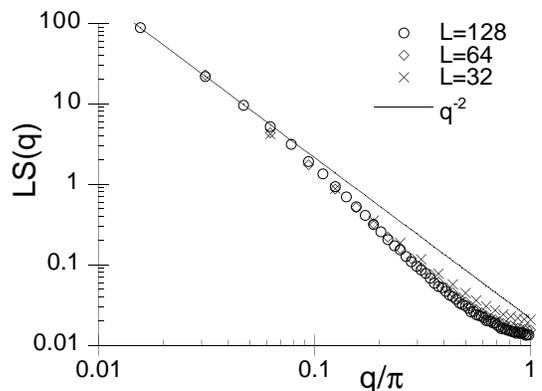}}
\caption{The static structure factor for interface fluctuations in a model
of MBE that is unstable toward mound formation \protect\cite{SP96} for
substrate sizes $L_x=L_y=$ 32, 64, 96 obtained from Monte Carlo simulations.
Parameters are as given in \protect\cite{SP96}. Note that the data collapse
only if $S(q)$ is multiplied by the substrate size $L$.}
\label{figsk}
\end{figure}

We now discuss numerical calculations that verify the properties derived
above. We have integrated Eq.~(\ref{current}) and also carried out Monte
Carlo simulations for a discrete growth model \cite{SP96}. Both display the
same kind of instability and coarsening behavior. In the latter, particles
are randomly deposited on an initially flat substrate and the growing
cluster relaxes by surface diffusion. An instability is produced by
step-edge barriers in one of the two hopping directions on the square
lattice. The steady state for a finite system is then a single mound with
its ridge and valley perpendicular to this axis. In Fig.~\ref{figsk} we show
the static structure factor $S_I(q)$ for the ridge for three different
substrate sizes with $L_x=L_y=L$. The collapse of the data, when $S_I(q)$ is
multiplied by $L$ is quite impressive. Further, the data seem to be converging
to the $q^{-2}$-dependence predicted above.

\begin{figure}[t]
\centerline{\epsfxsize=85mm\epsfbox[93 342 577 655]{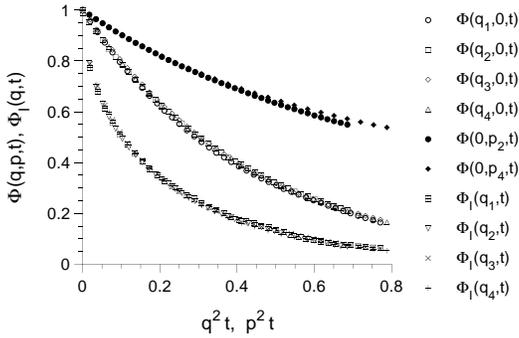}}
\caption{Relaxation functions $\Phi(q,p,t)$ and $\Phi_I(q,t)$ obtained from
a numerical integration of Eq.~(\ref{current}) as a function of the scaled
arguments $q^2t$ and $p^2t$, $q_n=2\pi n/L_x$ and $p_n=2\pi n/L_y$,
$L_x=L_y=192$.}
\label{c(qt)}
\end{figure}

In Fig.~\ref{c(qt)} we show the relaxation functions 
$\Phi(q,p,t)\equiv S(q,p,t)/S(q,p,0)$ and $\Phi_I(q,t)\equiv S_I(q,t)/S_I(q,0)$
obtained by integration of Eq.~(\ref{current}). The data collapse perfectly to
single curves when plotted as a function of $q^2t$, $p^2t$ respectively, in
complete agreement with (\ref{Skt}) and (\ref{Si_res}). The only exceptions
are the modes $S(0,p_n,t)$ with $p_n=2\pi n/L_y$ and $n$ an odd integer. For
these modes, the difference of two large quantities is encountered and we
have not succeeded in obtaining the necessary statistical and numerical
accuracy to reach a definitive conclusion. More details of our study,
including tests for the $L_y$ dependence using rectangular lattices, will be
presented elsewhere\cite{ZSP}.

In summary, we have demonstrated both by analytic calculations and by
simulations that this class of nonequilibrium processes behaves quite
differently from models A and B. In particular, the exponent $z$ that
controls the relaxation of interface fluctuations in the steady state is 
{\em not} related to the coarsening exponent $n$ by $n=1/z$. While this is
interesting in its own right, it also indicates that a theory of coarsening
for these processes must be based on the interactions of more complicated
defects than those allowed in the steady state. However, if the exponent
relation $\alpha'=2n$\cite{RK} is used, our result $z=2=1/\alpha'$ yields a
coarsening exponent $n=1/4$.

The results derived in this letter are also relevant for the interpretation
of experimental results obtained, e.g., in scattering experiments of films
grown using MBE. If the surface evolution can be described by pyramid
formation and coarsening, one expects that the scattering amplitude has a
``ring'' structure\cite{note}, i.e., a maximum at finite wavevectors 
${\bf k}_{{\rm max}}(t)\sim 1/R(t)\sim t^{-n}$. However, once $k_{\rm max}$ 
becomes smaller than the minimum resolvable wavelength, the observed
scattering is due to fluctuations of a single pyramid. In that case, the
dynamic exponent $z=2$ as explained above. Moreover, the measured
roughness exponent $\alpha$ then also characterizes fluctuations
superimposed on the pyramid structure and has the value $\alpha=0$.
Experimentally, it may be very difficult to detect this
logarithmic roughness.

This research was supported by the NSERC of Canada and the US National
Science Foundation through the Division of Materials Research. One of us
(RKPZ) is grateful for the hospitality of M. Wortis and the physics
department of Simon Fraser University.


\begin{references}
\bibitem{note}The word ``ring'' should not be taken literally, because the
anisotropies of the lattice structure will also be seen in the ring
structure.
\end{references}
\end{document}